\documentclass[aip,graphicx, amsmath, amssymb]{revtex4-1}
\usepackage{graphicx}
\usepackage{color}
\DeclareMathOperator{\Tr}{Tr}
\draft 

\newcommand{\COMMENTED}[1]{}
\newcommand{\REMARKS}[1]
{
{ \color{red}{\textbf{ {[#1]} }} }
}

\newcommand{\HS}[1]
{
{ \color{blue}{\textbf{ {[#1]} }} }
}
\begin{document}


\title{Some Recent Developments in 
Auxiliary-Field Quantum Monte Carlo for Real Materials} 



\author{Hao Shi}
\affiliation{Center for Computational Quantum Physics, Flatiron Institute, New York, NY 10010, USA}
\affiliation{Department of Physics and Astronomy, University of Delaware, Delaware, 19716, USA}
\author{Shiwei Zhang}
\affiliation{Center for Computational Quantum Physics, Flatiron Institute, New York, NY 10010, USA}
\affiliation{Department of Physics, College of William and Mary, Williamsburg, Virginia 23187, USA}

\date{\today}

\begin{abstract}
The auxiliary-field quantum Monte Carlo (AFQMC) method is a 
general numerical method for 
correlated many-electron systems, which is 
being increasingly 
applied in lattice models, atoms, molecules, and solids. Here we introduce the theory and algorithm of the method specialized for real materials, and present several recent developments. 
We give a systematic exposition of the key steps of AFQMC, 
closely tracking the framework of a modern software library we are developing.
The building of a Monte Carlo Hamiltonian, projecting to the ground state, sampling two-body operators, phaseless approximation, and measuring ground state properties are discussed in details. An advanced implementation for multi-determinant trial wave functions is described which dramatically speeds up the algorithm and reduces the memory cost. We propose a self-consistent constraint for real materials, and discuss two flavors for its realization, either by
 coupling the AFQMC calculation to an effective independent-electron calculation, or via the natural orbitals of the computed one-body density matrix.
\end{abstract}


\maketitle 

\section{Introduction}
The quantum many-body problem is one of the most challenging problems in the fields of condensed matter physics, quantum chemistry, and materials science. The properties of these systems result from the competition between the atomic environment, quantum delocalization of electrons, and electron-electron interaction. Accurate and reliable 
computations are essential for understanding and predicting the materials properties. The cost 
of finding the exact properties of these systems generally grows exponentially with the number of electrons and size of the systems, which motivates the development of modern numerical methods capable of 
approximate but sufficiently and systematically accurate solutions.

The most widely used numerical approaches for the many-body Schrodinger equation are based on density-functional theory (DFT) \cite{Martin2004, RevModPhys.71.1253}, which approximates the many-body effects by an auxiliary one-electron problem with an external potential. These approaches have been effective in most molecules and solids, and became the standard in 
electronic structure calculations. However, in the presence of strong electron-electron interaction, 
DFT-based approaches 
have not yet achieved the desired predictive power.
Systematic approaches beyond independent-particle theories are intensely investigated for strongly correlated systems.

Numerical methods using Monte Carlo (MC) sampling techniques \cite{PhysRevA.9.2178, RevModPhys.73.33, RevModPhys.67.279, PhysRevD.24.2278, PhysRevLett.90.136401} are promising in handling strongly-correlated electrons. These  MC methods 
allow non-perturbative treatments beyond DFT and tend to scale well (low-power) with system size. However, 
in general there is 
a ``sign" problem \cite{Schmidt1984,PhysRevB.41.9301, Zhang1999ConstrainedPM,Zhang_Book_2013} for fermion systems, which arises from negative signs in the wave function 
under the interchange of two fermions. The sign problem manifests itself as the cancellation among the  contributions of different MC samples, which 
becomes more severe as the system size is increased. In some cases, a phase problem appears which leads to more severe cancellations. Such cancellations cause the MC signal to decay exponentially versus noise. The unsolved sign problem and phase problem hinder the studies of the physics of many-fermion 
systems.

In this article, we discuss the auxiliary-field quantum Monte Carlo (AFQMC) method \cite{PhysRevLett.90.136401, doi:10.1063/1.2200885, doi:10.1002/wcms.1364, Zhang_Book_2013, PhysRevB.101.235110} for real materials. 
This method controls the sign and phase problem by a constraint in 
the sign or gauge of path integrals in auxiliary-field space.
It has demonstrated excellent versatility and accuracy across a wide range of systems and, in addition to lattice models, is being increasingly applied in molecules and solids.
We present a self-contained description of the method, providing a more unified 
framework between the formalisms for lattice models (short-range interaction),
periodic solids with plane-waves, and molecular systems using quantum chemistry 
machinery. 
This formalism emphasizes a generic Hamiltonian form for AFQMC, and a treatment of 
the Hubbard-Stratonovich transformation that does not distinguish between discrete 
and continuous fields. It 
parallels a modern software library that we are 
developing which spans the multiple application domains. 
We discuss our advanced implementations 
to speed up the simulations and reduce the computational time, especially for
 multi-determinant trial wave functions. A self-consistent  method
is introduced which couples the AFQMC calculation to an independent-electron calculation  to systematically improve the constraint. 

\COMMENTED{
In AFQMC, the interaction part of the Hamiltonian is cast into a summation of non-interacting terms through a Hubbard-Stratonovich transformation. The summation is then sampled by random walks in the space of over-complete Slater determinants, which significantly reduces the sign problem.  We introduce the constrained path \cite{PhysRevB.55.7464} and phaseless approximation \cite{PhysRevLett.90.136401}, which control the sign and phase problems to restore low-polynomial scaling,
at the cost of losing exactness
in the simulation. }

\section{Method}

In this section we give a self-contained description of the AFQMC method.
In AFQMC, the interaction part of the Hamiltonian is cast into a summation of non-interacting terms through a Hubbard-Stratonovich transformation. This is 
discussed in the form of a generic ``Monte Carlo Hamiltonian" in the first 
and second parts.
The summation is then sampled by random walks in the space of over-complete Slater determinants, which significantly reduces the sign problem. 
The projection and the random walk are outlined in the second and third parts, 
emphasizing the importance-sampling transformation in the language of a force
bias  which is independent of the detailed form of the auxiliary-fields (discrete 
or continuous). 
We then introduce the constrained path \cite{PhysRevB.55.7464} and phaseless \cite{PhysRevLett.90.136401} approximation, which control the sign and phase problems to restore low-polynomial scaling,
at the cost of losing exactness
in the simulation. 

\subsection{Hamiltonian}

The constrained-path and phaseless AFQMC methods have 
often been formulated in different 
flavors, depending on their target applications.
These include, in addition to lattice models \cite{PhysRevB.55.7464,PhysRevB.99.235142} and
Hartree-Fock-Bogoliubov approach \cite{PhysRevB.95.045144},
several variants for realistic systems distinguished by the
basis sets employed:
solids with plane-wave basis \cite{PhysRevLett.90.136401,Suewattana_PRB75_2007},
molecules with standard quantum chemistry basis sets of
Gaussian type orbitals (GTOs) \cite{doi:10.1063/1.2200885,doi:10.1002/wcms.1364}, 
and downfolded Hamiltonians \cite{Ma_PRL114_2015}
which treats solids but uses more of the GTO flavor of 
AFQMC.
Here we seek to better unify the different flavors 
through 
a common starting point of an AFQMC calculation. 
The two forms of the Hamiltonian separately discussed below are the same, but re-expressing
the original ``\textit{ab initio} Hamiltonian'' into a
``Monte Carlo Hamiltonian" 
allows us to introduce a more general 
discussion of the AFQMC
algorithm and a more uniform framework to think about 
the algorithm and implementation.

\subsubsection{\textit{Ab initio} Hamiltonian}
We start from the Born–Oppenheimer approximation \cite{doi:10.1002/andp.19273892002, Szabo_book} for real materials. The first quantized Hamiltonian for many-electron systems is
\begin{equation}
\hat{H}=\sum_{i=1}^{N}\left[-\frac{1}{2}\nabla_{i}^{2}+V_{ext}(r_{i})\right]+\sum_{i<j}^{N}\frac{1}{r_{ij}}\,,    
\end{equation}
where $r_i$ is the 3-dimensional coordinates of electron $i$, $r_{ij}=|r_i-r_j|$, and we have used atomic units.

The first part is the one-body term, which contains the kinetic energy and external potential, and the second part is the two-body term. Let us define $h(r)=-\frac{1}{2}\nabla^{2}+V_{ext}(r)$
and $V(r_{1},r_{2})=\frac{1}{r_{12}}$. By choosing the appropriate basis $\phi_i(r)$,
we can write the Hamiltonian into second-quantized form,
\begin{equation}
  \hat{H}=\sum_{ij}^{M} \sum_\sigma h_{ij}a_{i\sigma}^{\dagger}a_{j\sigma}+\frac{1}{2}\sum_{ijkl}^{M}\sum_{\sigma\rho}V_{ijkl}a_{i\sigma}^{\dagger}a_{j\rho}^{\dagger}a_{k\rho}a_{l\sigma}\,,
\end{equation}
with
\begin{equation}
    h_{ij}=\int dr\,\phi_{i}^{*}(r)h(r)\phi_{j}(r)\:,
\end{equation}
and
\begin{equation}
V_{ijkl}=\int\int dr_{1}dr_{2}\,\phi_{i}^{*}(r_{1})\phi_{j}^{*}(r_{2})V(r_{1},r_{2})\phi_{l}(r_{1})\phi_{k}(r_{2})\:.
\end{equation}
Here we use $M$ for the number of basis, $ijkl$ for the index of basis and $\sigma \rho$ for the index of spin. 
Note that $V_{ijkl}$ has $4$-fold symmetry, 
\begin{equation}
V_{ijkl}=V_{jilk}=V_{klij}^{*}=V_{lkji}^{*} \:.
\end{equation}
If the basis set is real, there will be the additional symmetry in $V_{ijkl}$, as is typically the case in 
quantum chemistry:
\begin{equation}
 V_{ijkl}=V_{ikjl}\, .   
\end{equation}

\subsubsection{Monte Carlo Hamiltonian}
Auxiliary-field quantum Monte Carlo can be used for any Hamiltonian as long as it can be written into the form of a \textit{Monte Carlo Hamiltonian}
\begin{equation}
    \hat{H}_{\rm mc} = \hat{T} + \frac{1}{2}\sum_\gamma^\Gamma \hat{L}^2_\gamma +C \:,
    \label{eq:omcHamiltonian}
\end{equation}
where $C$ is a real number, which is the nuclei repulsive energy 
for chemical systems.
Both $\hat{T}$ and $\hat{L}_\gamma$ are one-body operators, 
whose general form is
\begin{equation}
\hat{O}_1 = \sum_{ij}^M \sum_{\sigma} O^\sigma_{ij} a_{i\sigma}^{\dagger}a_{j\sigma}\:.
\end{equation}
Here $\hat{O}_1$ does not need to preserve symmetry between different spin components; it 
can be a color spin or a spinless operator for the corresponding systems.

There are a number of ways to transform an \textit{ab initio} Hamiltonian into a Monte Carlo Hamiltonian,
which can lead to AFQMC calculations of different efficiency 
or even systematic accuracy. 
The 4-rank tensor $V_{ijkl}$ can be written into a matrix format by grouping $n =(il)$ and $m=(kj)$ index, 
\begin{equation}
   V_{ijkl} = V_{(il),(kj)} = V_{nm}\,, 
\end{equation}
which is a Hermitian matrix
due to the symmetry in $V_{ijkl}$.
For the most general case, a straightforward approach is 
diagonalizing the $V_{mn}$ matrix,
\begin{align}
V_{mn} & =\sum_{\gamma}U_{m\gamma}D_{\gamma}U_{\gamma n}^{\dagger} \\
 & =\sum_{\gamma} \left(\sqrt{D_{\gamma}} U_{m\gamma}\right) \left( \sqrt{D_{\gamma}} U_{nr}^{*} \right) \\
  & =\sum_{\gamma}L_{m\gamma}L_{nr}^{*}\,. 
\end{align}
This has a high cost and is not practical for large $M$.

For quantum chemistry applications, the most common 
approach has been using 
modified Cholesky decomposition \cite{doi:10.1002/qua.560120408, doi:10.1063/1.1578621, doi:10.1002/jcc.21318,Purwanto_JCP135_2011}.
\COMMENTED{
The 4-rank tensor $V_{ijkl}$ can be written into a matrix format by grouping $n =(il)$ and $m=(kj)$ index, 
\begin{equation}
   V_{ijkl} = V_{(il),(kj)} = V_{nm}\,, 
\end{equation}
which is a Hermitian matrix
due to the symmetry in $V_{ijkl}$.
Due to the symmetry in $V_{ijkl}$, $V_{nm}$ is an Hermitian matrix and 
we can write 
\begin{equation}
    V_{mn}=\sum_{\gamma}^{\Gamma}L_{m\gamma}L_{nr}^{*}
\end{equation}
 directly with modified Cholesky decomposition \cite{doi:10.1002/qua.560120408, doi:10.1063/1.1578621, doi:10.1002/jcc.21318}.
}
Let us assume that we have already written the decomposition to the $J^{\rm th}$ step
\begin{align}
V_{mn} & =\sum_{\gamma}^{J}L_{m\gamma}L_{n\gamma}^{*}+\Delta_{mn}^{J}\:\\
 & =V_{mn}^{J}+\Delta_{mn}^{J}\:,
\end{align}
where $\Delta_{mn}^{J}$ is the reminder between $V_{mn}$ and $V_{mn}^{J}$.
We can generate the next Cholesky vector by
\begin{equation}
 L_{m(J+1)}=\frac{\Delta_{m[n]_{J}}^{J}}{\sqrt{\text{\ensuremath{\Delta_{[n]_{J}[n]_{J}}^{J}}}}}\:,   
\end{equation}
with $[n]_{J}$ the index of largest diagonal elements of $\Delta_{mn}^{J}$.
If $\ensuremath{\Delta_{[n]_{J}[n]_{J}}^{J}}\leq \epsilon$, then
all the elements 
\begin{equation}
    |V_{mn}-V_{mn}^{J}|=|\Delta_{mn}^{J}| \leq \epsilon\:.
\end{equation}
Empirically, we choose $\epsilon \sim 10^{-6}$ 
for quantum chemical systems and the total number of Cholesky vectors, $\Gamma$, is around $10M$ after the truncation.

With the modified Cholesky decomposition, we can 
turn the \textit{ab initio} Hamiltonian into
\begin{equation}
   H=\sum_{ij}^{M}\sum_{\sigma} h_{ij}a_{i\sigma}^{\dagger}a_{j\sigma}+\frac{1}{2}\sum_{\gamma}^{\Gamma}\sum_{ijkl}^{M}\sum_{\sigma\rho}L_{(il)\gamma}L_{(kj)\gamma}^{*}a_{i\sigma}^{\dagger}a_{j\rho}^{\dagger}a_{k\rho}a_{l\sigma}\,. 
\end{equation}
After regrouping the index, we have
\begin{align}
H & =\sum_{ij}^{M}\sum_{\sigma}h_{ij}a_{i\sigma}^{\dagger}a_{j\sigma}-\frac{1}{2}\sum_{\gamma}^{\Gamma}\sum_{ijkl}^{M}\sum_{\sigma\rho}L_{(il)\gamma}L_{(kj)\gamma}^{*}a_{i\sigma}^{\dagger}(\delta_{jl}\delta_{\rho\sigma}-a_{l\sigma}a_{j\rho}^{\dagger})a_{k\rho}\\
 & =\sum_{ij}^{M}\sum_{\sigma}\left[h_{ij}-\frac{1}{2}\sum_{\gamma}^{\Gamma}\sum_{k}^{M}L_{(ik)\gamma}L_{(jk)\gamma}^{*}\right]a_{i\sigma}^{\dagger}a_{j\sigma}  \nonumber \\
 & \; \;\;\;+\frac{1}{2}\sum_{\gamma}^{\Gamma}\left(\sum_{il}^{M}\sum_{\sigma}L_{(il)\gamma}a_{i\sigma}^{\dagger}a_{l\sigma}\right)\left(\sum_{jk}^{M}\sum_{\rho}L_{(kj)\gamma}^{*}a_{j\rho}^{\dagger}a_{k\rho}\right) \label{eq:mcHamiltonian}
\end{align}
Equation~(\ref{eq:mcHamiltonian}) is a Monte Carlo Hamiltonian, with
\begin{equation}
    \hat{T} = \sum_{ij}^{M}\sum_{\sigma}\left[h_{ij}-\frac{1}{2}\sum_{\gamma}^{\Gamma}\sum_{k}^{M}L_{(ik)\gamma}L_{(jk)\gamma}^{*}\right]a_{i\sigma}^{\dagger}a_{j\sigma} \: ,
\end{equation}
and 
\begin{equation}
    \hat{L}_\gamma = \sum_{il}^{M}\sum_{\sigma}L_{(il)\gamma}a_{i\sigma}^{\dagger}a_{l\sigma} \: .
\end{equation}
Note that we have used the property that $L_{(il)\gamma} = L^*_{(li)\gamma} $ for systems with real basis functions.

\COMMENTED{
There are many ways other than the modified Cholesky decomposition to transfer the \textit{ab initio} Hamiltonian to the Monte Carlo Hamiltonian, for example, diagonalizing the $V_{mn}$ matrix,
\begin{align}
V_{mn} & =\sum_{\gamma}U_{m\gamma}D_{\gamma}U_{\gamma n}^{\dagger} \\
 & =\sum_{\gamma} \left(\sqrt{D_{\gamma}} U_{m\gamma}\right) \left( \sqrt{D_{\gamma}} U_{nr}^{*} \right) \\
  & =\sum_{\gamma}L_{m\gamma}L_{nr}^{*} 
\end{align}
and low rank tensor decomposition.
}

Recently density-fitting \cite{SheeGPU} and low-rank tensor decomposition \cite{doi:10.1063/5.0004860,doi:10.1021/acs.jctc.8b00996}
have also been adopted. For plane-wave calculations, the 
Coulomb repulsion is naturally bilinear in momentum space
which can be decomposed analytically \cite{PhysRevLett.90.136401,Suewattana_PRB75_2007}. 
And of course in lattice models with short-range interactions
specialized decompositions can be used \cite{PhysRevB.55.7464,PhysRevB.99.235142}.
All of these forms can be cast in the form 
of the Monte Carlo Hamiltonian, which we will use 
as the starting point of the AFQMC calculations below.


\subsection{Projection}
AFQMC solves the ground state Schrodinger equation by the imaginary-time projection
\begin{equation}
|\Psi_{0} \rangle \propto \lim_{\beta \rightarrow \infty} \left(e^{-\beta \hat{H}_{\rm mc}}\right) | \Psi_{I} \rangle. 
\end{equation}
As long as the initial wave function is not orthogonal to the ground state wave function ($\langle \Psi_0| \Psi_{I} \rangle \neq 0 $), it will converge to the ground state when the imaginary projection time $\beta$ approaches infinite. In practice, $\beta$ is discretized into $n$ small time slices, with a time step $\Delta \tau = \beta / n$. The projection can be evaluated as
\begin{equation}
|\Psi_{0} \rangle \propto \lim_{n \rightarrow \infty} \left(e^{-\Delta \tau \hat{H}_{\rm mc}}\right) ^{n} | \Psi_{I} \rangle. 
\label{eq:project}
\end{equation}

For sufficiently small time step, the projection operator of the Monte Carlo Hamiltonian in Eq.~(\ref{eq:omcHamiltonian}) can be factorized into one-body and two-body parts by Suzuki-Trotter decomposition \cite{Suzuki_ProgTheorPhys_1976,Trotter_PAMS_1959}
\begin{align}
    e^{-\Delta\tau \hat{H}_{mc}} = e^{-\Delta\tau \hat{T}/2} e^{-\Delta\tau \frac{1}{2} \sum_\gamma\hat{L}^2_\gamma} e^{-\Delta\tau \hat{T}/2} e^{-\Delta \tau C} + \mathcal{O}\left(\Delta \tau^3\right).
\label{Projection_Equation}
\end{align}
The two-body propagators can be decomposed into one-body propagators by Hubbard-Stratonovich transformation  \cite{HS_PRL_1959}
\begin{equation}
e^{-\Delta \tau \hat{O}_1^{2}/2} = \int dx \frac{1}{\sqrt{2\pi}} e^{-x^{2}/2} e^{x\sqrt{-\Delta \tau}\hat{O}_1}
\label{continuous_HS_transformation}, 
\end{equation}
where $\hat{O}_1$ represents a one-body operator and $x$ 
is an auxiliary field.
The projection operator becomes the integration of one-body operators in a high-dimensional auxiliary-field space  
\begin{equation}
 e^{-\Delta\tau \hat{H}_{\rm mc}} = \int \prod_\gamma \left( dx_\gamma \frac{1}{\sqrt{2\pi}}  e^{-x_\gamma^{2}/2} \right)  e^{-\Delta\tau \hat{T}/2} 
 e^{\sum_\gamma x_\gamma\sqrt{-\Delta \tau}\hat{L}_\gamma}
 e^{-\Delta\tau \hat{T}/2} e^{-\Delta \tau C} + \mathcal{O}\left(\Delta \tau^2\right).
\end{equation}
Note that the Trotter error increase to $\mathcal{O}\left(\Delta \tau^2\right) $ during the process of grouping $\hat{L}_\gamma$ operators. The final expression of the projection operator is
\begin{equation}
e^{-\Delta \tau \hat{H}_{\rm mc}} = \int d{\bf x} p({\bf x}) \hat{B}({\bf x}),
\label{eq:effective_propagation}
\end{equation}
where ${\bf x} = \{x_1, x_2, . . . , x_{\Gamma}
\}$ denotes the 
auxiliary-field variables 
at a given time slice, $p({\bf x})$ is the probability function
\begin{equation}
    p({\bf x}) = \prod_\gamma \frac{1}{\sqrt{2\pi}}  e^{-x_\gamma^{2}/2} \: ,
\end{equation}
and $\hat{B}({\bf x})$ is the combination of all one-body operators
\begin{equation}
    \hat{B}({\bf x}) = e^{-\Delta\tau \hat{T}/2} 
 e^{\sum_\gamma x_\gamma\sqrt{-\Delta \tau}\hat{L}_\gamma}
 e^{-\Delta\tau \hat{T}/2} + \mathcal{O}\left(\Delta \tau^2\right).
\end{equation}
The original projector is mapped into a high-dimensional integral of auxiliary-fields over one-body propagators, which can be evaluated by Monte Carlo techniques.

As we can see, the two-body propagator $ e^{-\Delta\tau \frac{1}{2} \sum_\gamma\hat{L}^2_\gamma}$ is the one that generates the high-dimensional integral and it is more computational expensive than the one-body term $e^{-\Delta\tau \hat{T}/2}$. The algorithm will be more efficient if the magnitude of the two-body term is reduced. We can change the Monte Carlo Hamiltonian with a background subtraction trick \cite{Purwanto_PRA_2005, PhysRevB.99.235142},
\begin{equation}
    \hat{H}_{\rm mc} = \hat{T}+\sum_\gamma^\Gamma\langle \hat{L}_{\gamma} \rangle \hat{L}_\gamma + \frac{1}{2}\sum_\gamma^\Gamma \left( \hat{L}_\gamma -\langle \hat{L}_{\gamma} \rangle \right )^2 + C- \frac{1}{2}\sum_\gamma^\Gamma \langle \hat{L}_{\gamma} \rangle ^2 \:,
    \label{eq:omcHamiltonian-bg}
\end{equation}
which is still a Monte Carlo Hamiltonian with
\begin{align*}
     \hat{T} &\leftarrow \hat{T}+\sum_\gamma^\Gamma\langle \hat{L}_{\gamma} \rangle ,\\
    \hat{L}_\gamma&\leftarrow \hat{L}_\gamma -\langle \hat{L}_{\gamma} \rangle ,  \\
    C &\leftarrow C- \frac{1}{2}\sum_\gamma^\Gamma \langle \hat{L}_{\gamma} \rangle ^2 .
\end{align*}
The background subtraction applies to both electronic and lattice Hamiltonians, i.e., 
regardless of the details of the interaction or the form of the Hubbard-Stratonovich transformation. 

\subsection{Sampling}
Monte Carlo technique is one of the most efficient methods to calculate the high-dimensional integration. 
Early formulation of auxiliary-field-based methods \cite{PhysRevD.24.2278, SUGIYAMA19861,PhysRevB.31.4403, Sorella_1989,PhysRevB.40.506, PhysRevE.93.033303} 
was based on the Metropolis algorithm, which is highly 
effective when there is no sign/phase problems 
(however care should be taken in handling an infinite variance problem \cite{PhysRevE.93.033303}). 
In the presence of the sign/phase problem, which is the case with all 
quantum chemical systems and all realistic materials computations, 
the reformulation of the framework into an 
open-ended random walk \cite{PhysRevB.55.7464} was essential.
The open-ended random walk removes an ergodicity problem in 
the path-integral formulation which renders the sampling of positive 
paths exponentially-costly in at low-temperatures \cite{PhysRevB.55.7464,Zhang1999ConstrainedPM}. 
Moreover, it provides a conceptual framework \cite{PhysRevLett.90.136401,Zhang_Book_2019} closely 
aligned with standard DFT machinery, which allowed successful and  general application to electronic structure.

\subsubsection{Free Projection}
We initialize the $| \Psi_{I} \rangle$ to a Slater determinant, which is usually the Hartree-Fock (HF) solution $| \psi_{\rm HF}\rangle$. The initial many-body wave function can be thought of as a summation of Slater determinants (so-called walkers)
\begin{equation}
    | \Psi_{I} \rangle = \sum_k w^{(0)}_k | \psi_k^{(0)} \rangle \:,
\end{equation}
where $w^{(0)}_k=1$ and $| \psi_k^{(0)}\rangle = | \psi_{\rm HF}\rangle $. The number of walkers is usually between a few hundreds to a few thousands, but can be tuned according to the computing platform to maximize efficiency for improving statistics. The projection is applied to the initial state as shown in Eq.~(\ref{eq:project}), and the overall wave function after $n$ steps can
be schematically represented as
\begin{equation}
    | \Psi^{(n)} \rangle = \sum_k w^{(n)}_k | \psi_k^{(n)} \rangle \:.
\label{eq:wf-MC}
\end{equation}
The weight $w^{(n)}_k$ contains products of numbers accumulated during the projection, which can be a complex number, and $| \psi_k^{(n)} \rangle $ is still a Slater determinant. The projection to the next step is carried out by
\begin{equation}
     | \Psi^{(n+1)} \rangle =  e^{-\Delta\tau \hat{H}_{mc}} | \Psi^{(n)} \rangle \:.
\end{equation}
Using  Eq.~(\ref{eq:effective_propagation}), we obtain the auxiliary field $\bf x_k$ by sampling $p({\bf x_k})$ and applying the one-body operator to the walker $k$,
\begin{equation}
    | \psi_k^{(n+1)} \rangle = \hat{B}({\bf x_k}) | \psi_k^{(n)}  \rangle \:.
\end{equation}
Note that the new walker remains 
a Slater determinant due to the Thouless theorem \cite{THOULESS1960225, THOULESS196178}. All the numbers during the projection are absorbed into $w^{(n)}_k$ to produce the new $w^{(n+1)}_k$. To keep the walker 
numerically stable during the propagation, the modified Gram-Schmidt procedure is applied to each walker, and the normalization factor during the stabilization is absorbed into $w^{(n+1)}_k$.
As the imaginary time step increases, some walkers will contribute significantly more than other walkers. A population control procedure is needed to replicate the walkers with larger weights and eliminate  the walkers with smaller weights. There are different ways to choose the weights; for example, the weights in the algorithm without any importance sampling (free propagation) can be chosen as \cite{PhysRevB.88.125132}
\begin{equation}
W^{(n)}_{k} =  \sqrt{ |w_{k}^{(n)}|^2 \langle \psi_{k}^{(n)} | \psi_{k}^{(n)} \rangle } .
\end{equation}

\subsubsection{Importance Sampling}
We introduced the open-ended random walk procedure in the previous subsection, where the auxiliary fields are sampled by the function $p({\bf x_k})$. To further reduce the variance in the quantum Monte Carlo, we need to use importance sampling \cite{PhysRevB.55.7464,PhysRevLett.90.136401, PhysRevB.99.235142}. The method requires a best guess of the ground state wave function, which is called trial wave function, $| \Psi_T \rangle$. The trial wave function can be a 
HF or DFT solution, a multi-determinant wave function \cite{PhysRevB.88.125132, PhysRevB.89.125129}, or a Bardeen–Cooper–Schrieffer (BCS) wave function \cite{PhysRevA.84.061602, PhysRevA.92.033603, PhysRevB.95.045144}. 
The idea of the importance sampling transformation is 
to guide the random walks (the sampling of the auxiliary-fields and hence the resulting determinants) during the propagation 
towards regions with larger overlap with the trial wave function. 
The weights in population control with importance sampling are given by
\begin{equation}
W^{(n)}_{k} = w_{k}^{(n)} \langle \Psi_{T} | \psi_{k}^{(n)} \rangle\,.
\label{eq:weight}
\end{equation}
Note that $W^{(n)}_{k}$ can be a negative or complex number. When the constrained path or phaseless approximation is applied, $W^{(n)}_{k}$ is always positive or zero.

For sampling the auxiliary field, we change the probability density function from the product of Gaussians $p({\bf x_k})$ to a function that builds in the knowledge of 
$| \Psi_T \rangle$. The Hubbard-Stratonovich transformation in Eq.~(\ref{continuous_HS_transformation}) can be rewritten as 
\begin{equation}
\begin{split}
e^{-\Delta \tau \hat{O}_1^{2}/2} 
&= \int dx \frac{1}{\sqrt{2\pi}} e^{-x^{2}/2} e^{x\sqrt{-\Delta \tau}\hat{O}_1} \\
&= \int dx \frac{1}{\sqrt{2\pi}} e^{-x^{2}/2} e^{x\sqrt{-\Delta \tau}\langle \hat{O}_1 \rangle} e^{x\sqrt{-\Delta \tau}(\hat{O}_1-\langle \hat{O}_1 \rangle)},  
\label{importance_sampling_continuous}
\end{split}
\end{equation}
where $\langle \hat{O}_1 \rangle$ is the mixed estimator of $\hat{O}_1$ defined for a particular walker
\begin{equation}
\langle \hat{O}_1 \rangle
= \frac{\langle \Psi_T | \hat{O}_1 | \psi_k^{(n)} \rangle }{\langle \Psi_T | \psi_k^{(n)} \rangle }. 
\end{equation}
Let us define the dynamic force as $F \equiv \sqrt{-\Delta \tau}\langle \hat{O}_1 \rangle$. Then Eq.~(\ref{importance_sampling_continuous}) can be written as
\begin{equation}
\begin{split}
e^{-\Delta \tau \hat{O}_1^{2}/2} 
&= \int dx \frac{1}{\sqrt{2\pi}} e^{-x^{2}/2} e^{xF} e^{x \left(\sqrt{-\Delta \tau}\hat{O}_1-F \right)} \\
&=\int dx \frac{1}{\sqrt{2\pi}} e^{-x^{2}/2} e^{xF} \left[ 1+ x\left(\sqrt{-\Delta \tau}\hat{O}_1-F\right) + \mathcal{O}\left(\Delta \tau\right) \right]
\label{importance_sampling_continuous-2}. 
\end{split}
\end{equation}
To favor the sampling of walkers with more expected contributions to the ground state, we wish to build in the knowledge of the overlap between the trial wave function and the walker  
\begin{align}
    \frac{\langle \Psi_T |e^{-\Delta \tau \hat{O}_1^{2}/2}|\psi_k^{(n)} \rangle}{\langle \Psi_T |\psi_k^{(n)} \rangle}  & = \int dx \frac{1}{\sqrt{2\pi}} e^{-x^{2}/2} e^{xF} + \mathcal{O}\left(\Delta \tau\right) \\
    & = \int dx \frac{1}{\sqrt{2\pi}} e^{-\left(x-F\right)^{2}/2} e^{F^2/2} + \mathcal{O}\left(\Delta \tau\right)
    \:.
\end{align}
It is clear that the most efficient way (up to the order of $\Delta \tau$) to generate 
the auxiliary fields is by sampling the modified probability function
\begin{equation}
    p_I\left(x\right) = \frac{1}{\sqrt{2\pi}} e^{-\left(x-F\right)^{2}/2}\,.
\end{equation}
With the modified 
probability density function, the Hubbard-Stratonovich transformation can be written as 
\begin{align}
e^{-\Delta \tau \hat{A}^{2}/2} 
&= \int dx \frac{1}{\sqrt{2\pi}} e^{-(x-F)^{2}/2} e^{\frac{1}{2}F^2-xF}e^{x\sqrt{-\Delta \tau}\hat{A}} \:,\\
&= \int dx \, p_I\left(x\right) N_I\left(x\right)e^{x\sqrt{-\Delta \tau}\hat{A}} \:,
\end{align}
with $N_I\left(x\right)= e^{\frac{1}{2}F^2-xF}$.
Combining the different auxiliary-field components,  
we can again write the projection operator into a high dimensional integral,
\begin{equation}
e^{-\Delta \tau \hat{H}_{mc}} = \int d{\bf x} \, p_I({\bf x}) \hat{B}_I({\bf x}),
\end{equation}
where $p_I({\bf x})$ is the force-bias-shifted probability density function
\begin{equation}
    p_I({\bf x}) = \prod_\gamma p_I\left(x_\gamma \right) \: .
    \label{eqn:importancepdf}
\end{equation}
By sampling the new probability function $p_I(\bf x)$, we apply the projection operator to each walker
\begin{equation}
   w_k^{(n+1)} | \psi_k^{(n+1)} \rangle = \hat{B}_I({\bf x_k})  w_k^{(n)} | \psi_k^{(n)}  \rangle \:.
   \label{eqn:project}
\end{equation}
Note that the normalization factor $N_I\left(x\right)$ in $\hat{B}_I({\bf x_k})$ will be absorbed into $w_k^{(n)}$ and  the operators in $\hat{B}_I({\bf x_k})$ will be applied to $| \psi_k^{(n)}  \rangle $ during the propagation.

\COMMENTED{
The importance sampling uses the knowledge of the trial wave function $|\Psi_T \rangle$ to reduce the Monte Carlo fluctuation. In the above we have presented it as 
a reformation of the Hubbard-Stratonovich transformation
which does not affect the expected value of the Monte Carlo nor the convergence
time in a ground-state calculation (but the choice of the initial wavefunction or
population of walkers clearly can affect the convergence time). Its 
and speedup the convergence time. However, it does not introduce any bias in theory since it is only a reformation of Hubbard-Stratonovich transformation.
}

The importance sampling uses the knowledge of the trial wave function $|\Psi_T \rangle$ to reduce the Monte Carlo fluctuation. In the above we have presented it as a reformulation of the Hubbard-Stratonovich transformation.
Importance sampling in the usual sense does not change the 
expectation value, only the variance. When there is a 
sign problem, this holds in that 
the computed energy is always
the constrained-path result with the same trial 
wave function, whether importance sampling is applied or not. When there is a phase problem, however, 
we are using the ``importance-sampling'' transformation in 
\emph{an unconventional sense}, to select 
a unique 
gauge choice. In this case, the phaseless approximation is defined with respect to the trial wave 
function {\it after} importance sampling; the 
similarity transformation is essential and affects the 
expectation value as well as the variance
 \cite{PhysRevLett.90.136401}.
 
\subsubsection{Constrained Path and Phaseless Approximations}
We address the sign and phase problem in this subsection, which stems from the fact that a Slater determinant, $|\psi_k\rangle$, remains invariant to arbitrary rotations, such as $e^{i\theta} | \psi_k \rangle$. During the propagation, the walkers 
will have random contributions to the 
phase accumulated from the propagator $\hat{B}(\bf x)$, which contains stochastically sampled auxiliary-fields and complex matrix elements. These phases will 
eventually cause the random walkers to populate the entire complex plane and the statistical average of the walkers, 
in the sense of Eq.~(\ref{eq:wf-MC}),
will approach zero, which leads to an decay in observable signal-to-noise ratios. 
Unless the development of the phase (or sign) is prevented explicitly by symmetry, this decay will occur, and its onset is exponential
with projection time or inverse temperature \cite{PhysRevB.99.045108}.
The loss of signal manifests itself as infinite variance in the Monte Carlo estimators. 
(Note that the reverse is not true: an infinite variance problem can appear in a large number of sign-problem-free AFQMC calculations, requiring care to mitigate
\cite{PhysRevE.93.033303}.)


When the propagators $\hat{B}(\bf x)$ are real, the only possible phases in a walker are $0$ and $\pi$.
This is the commonly referred to 
sign problem, which occurs widely in lattice model calculations. We impose a constrained path approximation \cite{PhysRevB.55.7464,Zhang_Book_2019} by requiring that all walkers maintain a positive overlap with the trial wave function during the propagation
\begin{equation}
W_k^{(n)} >0 , 
\end{equation}
where $W_k^{(n)}$ is defined in Eq.~(\ref{eq:weight}). 
It can be shown that the constraint will be exact if the trail wave function is the ground state wave function. However, the ground state wave function is usually unknown, and the use of $|\Psi_T\rangle$ in
implementing the constraint will result in a small bias \cite{PhysRevX.5.041041, Zheng1155}.

The phaselsess approximation \cite{PhysRevLett.90.136401,Zhang_Book_2019} is employed to control the phase problem when the propagators are complex. 
It defines a unique gauge $\theta$ with respect to the knowledge of the true ground state, by
projecting a complex walker onto the positive real axis 
\begin{equation}
\label{cosProjection}
W^{(n)}_{k} \rightarrow {\rm Re}\left[W^{(n)}_{k}\right] \times \max (0, \cos(\Delta \theta)),
\end{equation}
where the phase angle $\Delta \theta$ for each component of the auxiliary-field $x$ is 
\begin{equation}
\Delta \theta ={\rm arg} \left[ \frac{\langle \Psi_T | \hat{B}(x) | \psi_k^{(n)} \rangle }{\langle \Psi_T | \psi_k^{(n)} \rangle } \right] \approx \mathcal{O}({\rm Im}(xF)).
\end{equation}
The cosine projection will ensure that the density of the walkers vanish at the origin of the complex plane of $\langle\Psi_0|\psi\rangle$ (or of the proxy 
$\langle\Psi_T|\psi\rangle$ when implemented as a phaseless approximation using $\Psi_T\rangle$).
Note that phaseless approximation is smoothly connected to 
the constrained path approximation, since $\Delta \theta$ is  zero if $\hat{B}(\bf x)$ is real \cite{PhysRevB.99.235142}.

\subsection{Measurement}
After the imaginary-time projection is converged within the expected statistical accuracy, we can measure the ground-state properties 
with  additional projection time
steps.
For a physical quantity $\hat{A}$ that commutes with the Hamiltonian, $[\hat{A}, \hat{H}] =0$, we use the mixed estimator
\begin{align}
  \langle \hat{A} \rangle_{mix} 
  &= \frac{\langle \Psi_{T} | \hat{A} | \Psi^{(n)} \rangle}{\langle \Psi_{T} | \Psi^{(n)} \rangle} \nonumber \\
&= \frac{ \sum_{k} w_{k}^{(n)} \langle \Psi_{T} | \hat{A} | \psi_{k}^{(n)} \rangle}{ \sum_{k} w_{k}^{(n)} \langle \Psi_{T} | \psi_{k}^{(n)} \rangle}\,.
\label{eq:Mixed}
\end{align}
It is common to introduce the ``local'' measurement: 
\begin{equation}
A_{L}[\Psi_T, \Phi] \equiv \frac{\langle \Psi_{T}| \hat{A} | \Phi \rangle}{\langle \Psi_{T} | \Phi \rangle}, 
\end{equation}
so that Eq.~(\ref{eq:Mixed}) can be easily calculated by
\begin{equation}
\langle \hat{A} \rangle_{mix} = \frac{ \sum_{k} W_{k}^{(n)} A_{L}[\Psi_T, \psi_{k}^{(n)}]}{ \sum_{k} W_{k}^{(n)} } \:. 
\label{Mixed_Again}
\end{equation}
Note that $W_{k}^{(n)}$ is the weight in Eq.~(\ref{eq:weight}), used in the population control following importance sampling transformation, 
both in the constrained path and phaseless formulations. 

If $\hat{A}$ does not commute with $\hat{H}$,
the mixed estimator is biased. The back-propagation technique \cite{PhysRevB.55.7464, PhysRevE.70.056702, doi:10.1021/acs.jctc.7b00730} was proposed to remove this bias, which measures the physical quantity by
\begin{equation}
  \langle \hat{A} \rangle_{bp} 
= \frac{ \sum_{k} w_{k}^{(n)} \langle \Psi_{T} |e^{-m\Delta\tau \hat{H}_{mc}}  \hat{A} | \psi_{k}^{(n)} \rangle}{ \sum_{k} w_{k}^{(n)} \langle \Psi_{T} |e^{-m\Delta\tau \hat{H}_{mc}} | \psi_{k}^{(n)} \rangle} \:.
\end{equation}
In the framework of open-ended random walkers, the back-propagation can be 
represented by
\begin{align}
  \langle \hat{A} \rangle_{bp} 
& = \frac{ \sum_{k} w_{k}^{(n+m)} \langle \Psi_{T} |\hat{B}({\bf x}_{n+m}) \cdots \hat{B}({\bf x}_{n+1})  \hat{A} | \psi_{k}^{(n)} \rangle}{ \sum_{k} w_{k}^{(n+m)} \langle \Psi_{T} | \psi_{k}^{(n+m)} \rangle} \\
& = \frac{ \sum_{k} W_{k}^{(n+m)} A_L[\phi_m, \Psi_k^{(n)}]}{ \sum_{k} W_{k}^{(n+m)} } \:,
\label{eq:back-propagation}
\end{align}
where the $\phi_m$ in the local measurement $A_L$ is the back-propagated Slater determinant
\begin{equation}
    |\phi_m\rangle = \hat{B}^\dagger({\bf x}_{n+1}) \cdots \hat{B}^\dagger({\bf x}_{n+m}) |\Psi_T \rangle \:.
\end{equation}
This formalism is exact and by-passes the difficulty of a brute-force estimator
of matching two independent populations for the bra and ket \cite{PhysRevE.70.056702}. 
In practice, the constrained path (phaseless) approximation breaks the symmetry in the imaginary time axis. The back-propagated wave function is not equal to $|\Psi^{CP}_0\rangle$ (or $|\Psi^{PL}_0\rangle$ in phaseless). This 
causes a bias which is typically larger than that of a pure 
estimator formed by two $|\Psi^{CP}_0\rangle$'s ($|\Psi^{PL}_0\rangle$'s).
In phaseless calculations, an additional improvement is achieved \cite{doi:10.1021/acs.jctc.7b00730} if 
we restore the phases to the weight $W_{k}^{(n+m)}$ (for the back-propagation portion only) in Eq.~(\ref{eq:back-propagation}),
\begin{equation}
    W_{k}^{(n+m)} = W_{k}^{(n+m)}\prod_{w=n+1}^{n+m}\frac{1}{\max \left[0, \cos(\Delta \theta_k^{(w)}) \right]} \:.
\end{equation}
Note that the back-propagation step $m$ needs to be large enough to reach the convergence of $|\phi_m \rangle$ while keeping 
the accumulated phase stable. 

\section{advanced implementation}
The AFQMC algorithm as outlined above
can be implemented by linear algebra operations for general basis sets. The scaling of the algorithm is $M^4$ in the naive implementation, 
with large prefactors. We discuss an advanced implementation in this section, which reduces the scaling to $M^2 N^2$ in the measurement. 
To facilitate the discussion, we sketch a summary of 
the AFQMC algorithm:
\begin{enumerate}
  \item Set up the initial state $| \Psi_{I} \rangle = \sum_k w_k | \psi_k \rangle$. For example we can choose $w_k=1$ and $| \psi_k \rangle$ as $|\Psi_T\rangle$.
  (A multi-determinant $\Psi_T\rangle$ can be sampled according to the 
  squared absolute value of the coefficients.)
  \item Compute 
  the overlap $\langle \Psi_{T} | \psi_{k} \rangle$  and apply the constrained path or phaselss approximation to the weight. The walker is killed by setting the weight to zero. 
  \item If the weight is non-zero, compute the dynamic force components \{$F_\gamma$\}.
  \item Sample the auxiliary field ${\bf x}$ from the modified probability function involving 
  the force bias components $F_\gamma$, $p_I({\bf x})$ in Eq~(\ref{eqn:importancepdf}). Calculate the phase $\Delta \theta_k$ for ${\rm cos}$ projection.
  \item Propagate the walker with $\hat B_I({\bf x})$, 
  and update the weight $ w_{k} $ according to the normalization (Eq~(\ref{eqn:project})).
 \item Repeat steps $2$ to $5$ for all walkers, which forms one step of the projection.
  \item Periodically perform population control procedure to adjust the weights
  $\{W_k\}$.
  \item Periodically perform  the modified Gram-Schmidt procedure to orthonormalize the orbitals of the walkers.
  \item Periodically measure the ground state properties after a sufficiently large  imaginary time of initial equilibration.
\end{enumerate}
There are three main computational components; 
\begin{itemize}
  \item Calculate the weight: $W_{k} = w_{k} \langle \Psi_{T} | \psi_{k} \rangle$
  \item Calculate the force bias: $F_\gamma = \sqrt{-\Delta \tau} \langle \Psi_T | \hat{L}_\gamma | \psi_k \rangle / \langle \Psi_T | \psi_k \rangle . $
  \item Compute the local observable in measurements $\langle \hat{A} \rangle  = \langle \Psi_T | \hat{A} | \psi_k \rangle / \langle \Psi_T | \psi_k \rangle . $
\end{itemize}

Below we separately discuss the details of how to 
speed up the calculations when the trial wave function is a \emph{single-determinant}
 and when it is a \emph{multi-determinant} in the form of the complete active-space self-consistent field (CASSCF).
 

\subsection{Single-determinant}
For a single-determinant trial wave function, we assume its matrix representation is $\left(\Psi_{T}^\uparrow, \Psi_{T}^\downarrow\right)$, where $\Psi_{T}^\sigma$ is a $M\times N^\sigma$ matrix and $N^\sigma$ is the number of particles for spin $\sigma$. The walkers in AFQMC are also single-determinants, each with the matrix representation $\left(\psi_{k}^\uparrow, \psi_{k}^\downarrow\right)$.

\subsubsection{Weight}
The weight $W_{k} = w_{k} \langle \Psi_{T} | \psi_{k} \rangle$ is calculated from 
the overlap between the trial wave function and each walker. In the matrix representation, the overlap is a determinant 
\begin{equation}
    \langle \Psi_{T} | \psi_{k} \rangle = \det \left( {\Psi_{T}^{\uparrow}}^\dagger \psi_k^{\uparrow} \right) \det \left( {\Psi_{T}^{\downarrow}}^\dagger \psi_k^{\downarrow} \right) .
\end{equation}
The computation of the 
overlap, with the scaling of ${MN^\sigma}^2+{N^\sigma}^3$, is generally small compared to other operations. 
We save the LU decomposition of the matrix $\Psi_{T}^{\sigma\dagger} \psi_k^{\sigma} $
when computing the overlap and determinant, 
which will be useful for calculating $\left(\Psi_{T}^{\sigma\dagger} \psi_k^{\sigma} \right)^{-1}$ and the Green's function in the following discussion.

\subsubsection{Force bias}
The force bias is the measurement of the one-body operator $\hat{L}_\gamma$. To measure any one-body operator, we first introduce the one-particle reduced density matrix
\begin{align}
    G^\sigma_{ij} &= \frac{\langle \Psi_T |a^\dagger_i a_j| \psi_k \rangle}{\langle \Psi_T | \psi_k \rangle}    \\
    &= \left[ \psi_k^{\sigma} \left(\Psi_{T}^{\sigma\dagger} \psi_k^{\sigma} \right)^{-1} \Psi_{T}^{\sigma\dagger} \right]_{ji} .
\end{align}
If the matrix representation of $\hat{L}_\gamma$ is $L^\sigma_\gamma$, the measurement of force is given by
\begin{align}
F_\gamma &= \sqrt{-\Delta \tau} \sum_{ij} \sum_\sigma \left(L^\sigma_{\gamma}\right)_{ij} G_{ij}^\sigma \\
    &= \sqrt{-\Delta \tau} \sum_\sigma \Tr \left[ L^\sigma_{\gamma} \psi_k^{\sigma} \left(\Psi_{T}^{\sigma}{}^\dagger \psi_k^{\sigma} \right)^{-1} \Psi_{T}^{\sigma}{}^\dagger \right] \\
    &= \sqrt{-\Delta \tau} \sum_\sigma \Tr \left[ \left( \Psi_{T}^{\sigma}{}^\dagger L^\sigma_{\gamma} \right) \Theta_k^\sigma \right], 
\end{align}
where $\Theta_k^\sigma = \psi_k^{\sigma} \left(\Psi_{T}^{\sigma}{}^\dagger \psi_k^{\sigma} \right)^{-1} $. Since $\Psi_{T}^{\sigma\dagger} L^\sigma_{\gamma}$ is independent of walkers during the propagation, we can pre-compute it 
and use the stored results throughout 
the whole AFQMC simulation. $\Theta_k^\sigma$ is calculated by  solving a linear equation with LU decomposition of ${\Psi_{T}^{\sigma}}^\dagger \psi_k^{\sigma} $ saved before. 
$\Theta_k^\sigma$ can be saved for 
the measurements of other quantities.  
With the procedures above, the calculation of the force bias only scales as $\Gamma M N^\sigma + M N^\sigma{}^2$.  

\subsubsection{Measurements}
For the measurement of any one-body operator, we use the same trick as in the calculation of the force bias. The most costly 
part in the measurement is the full Coulomb interaction 
energy
\begin{equation}
\hat{V}= \frac{1}{2}\sum_{\gamma}^{\Gamma}\sum_{ijkl}^{M}\sum_{\sigma\rho}L_{(il)\gamma}L_{(kj)\gamma}^{*} \frac{\langle \Psi_T |a_{i\sigma}^{\dagger}a_{j\rho}^{\dagger}a_{k\rho}a_{l\sigma}| \psi_k \rangle}{\langle \Psi_T | \psi_k \rangle} \, .
\end{equation}
With the generalized Wick’s theorem \cite{PhysRev.80.268, Balian:1969tb}, we have
\begin{equation}
    \frac{\langle \Psi_T |a_{i\sigma}^{\dagger}a_{j\rho}^{\dagger}a_{k\rho}a_{l\sigma}| \psi_k \rangle}{\langle \Psi_T | \psi_k \rangle}  = G_{il}^\sigma G_{jk}^\rho-\delta_{\sigma\rho} G_{ik}^\sigma G_{jl}^\sigma \, ,
\end{equation}
The interaction 
energy can be calculated as
\begin{align}
\hat{V} &= \frac{1}{2}\sum_{\gamma}^{\Gamma}\sum_{ijkl}^{M}\sum_{\sigma\rho}L_{(il)\gamma}L_{(kj)\gamma}^{*} \left( G_{il}^\sigma G_{jk}^\rho-\delta_{\sigma\rho} G_{ik}^\sigma G_{jl}^\sigma \right) \\
 &=  \frac{1}{2} \sum_{\gamma}^{\Gamma}\left[ \left(\frac{\langle \Psi_T | \hat{L}_\gamma | \psi_k \rangle}{ \langle \Psi_T | \psi_k \rangle }  \right)^2 - \sum_\sigma \Tr\left(  \Psi_{T}^{\sigma\dagger} L^\sigma_{\gamma} \Theta_k^\sigma \Psi_{T}^{\sigma\dagger} L^\sigma_{\gamma} \Theta_k^\sigma \right) \right],
\end{align}
where the first part (Hartree term) is just the square of the force. To handle the second part (exchange term), which is more costly, we order the matrix operations as 
\begin{equation}
    \left[\left(\Psi_{T}^{\sigma\dagger} L^\sigma_{\gamma} \right) \Theta_k^\sigma \right] \left[\left(\Psi_{T}^{\sigma\dagger} L^\sigma_{\gamma} \right) \Theta_k^\sigma \right],
\end{equation}
which scales as $\Gamma M N^{\sigma 2}$.
Recent progress has been able to reduce the scaling by introducing different forms of low-rank decomposition \cite{doi:10.1021/acs.jctc.8b00996,Lee-Reichman-RI-doi:10.1063/5.0015077,Malone-Tdoi:10.1021/acs.jctc.8b00944}. We will not discuss the implementation of these approaches here,
although they do not require 
fundamentally different computational ingredients beyond what we have covered. 

\subsection{Multi-determinant}
For a general multi-determinant trial wave function, we can apply the procedure 
discussed above for each determinant. This implementation will cause an additional scaling factor $N_d$, which is the number of determinants in the trial wave function. In the limit of large $N_d$, it becomes inefficient in both computational time and 
memory requirements. One of the most widely used trial wave functions is the CASSCF wave function, where 
all the determinants are built from the some canonical orbitals. It is natural to take advantage of the properties of the CASSCF wave function to reduce computational scaling, similar to 
how fast updates of one or a few components of the auxiliary-fields were 
handled via Sherman-Morrison formula. An implementation of the CASSCF trial wave function speedup was presented in Ref.~\cite{SheeGPU}. Here we present the details of our implementation which
reduces the scaling to sub-linear in $N_d$.

The CASSCF wave function has the form
\begin{equation}
    |\Psi_T\rangle = \sum_{m=1}^{N_d} c_m | \phi_m \rangle ,
\end{equation}
where the coefficient $c_m$ is a c-number and $| \phi_m \rangle $ is a Slater determinant with the matrix representation $\left(\phi_m^\uparrow, \phi_m^\downarrow\right)$. As mentioned before, pairs of these Slater determinants can share some columns with each other. It is also possible that the matrix for one spin component is identical in two determinants, 
for example, 
$\phi_m^\uparrow = \phi_n^\uparrow$, while $\phi_m^\downarrow \neq \phi_n^\downarrow$. 

We define the group 
$\varphi^\sigma = \left(\phi_1^\sigma, \phi_2^\sigma, \cdots, \phi_{S_d^\sigma}^\sigma \right)$, 
which removes duplicate Slater determinants, with 
$S_{d}^\sigma$ 
specifying the number of unique spin-$\sigma$ determinants. By the mapping from $\phi^\sigma $ to $\varphi^\sigma$, the CASSCF wave function becomes
\begin{equation}
    \label{eq:caswf}
    |\Psi_T\rangle = \sum_{m=1}^{N_d} c_m | \varphi_{m_\uparrow}^\uparrow, \varphi_{m_\downarrow}^\downarrow \rangle\,.
\end{equation}
Here, each $m$ is mapped to the index $(m_\uparrow, m_\downarrow)$ in the non-repetitive group $\varphi$. In the following, we only deal with $\varphi$, which reduces the number of operations from $N_d$ to 
$S_{d}^\uparrow + S_{d}^\downarrow$. Note that $S_{d}^\sigma \propto \sqrt{N_d}$ for the spin-balanced systems.

\begin{figure}[htbp]
\begin{center}
  \includegraphics[width=1.0\textwidth]{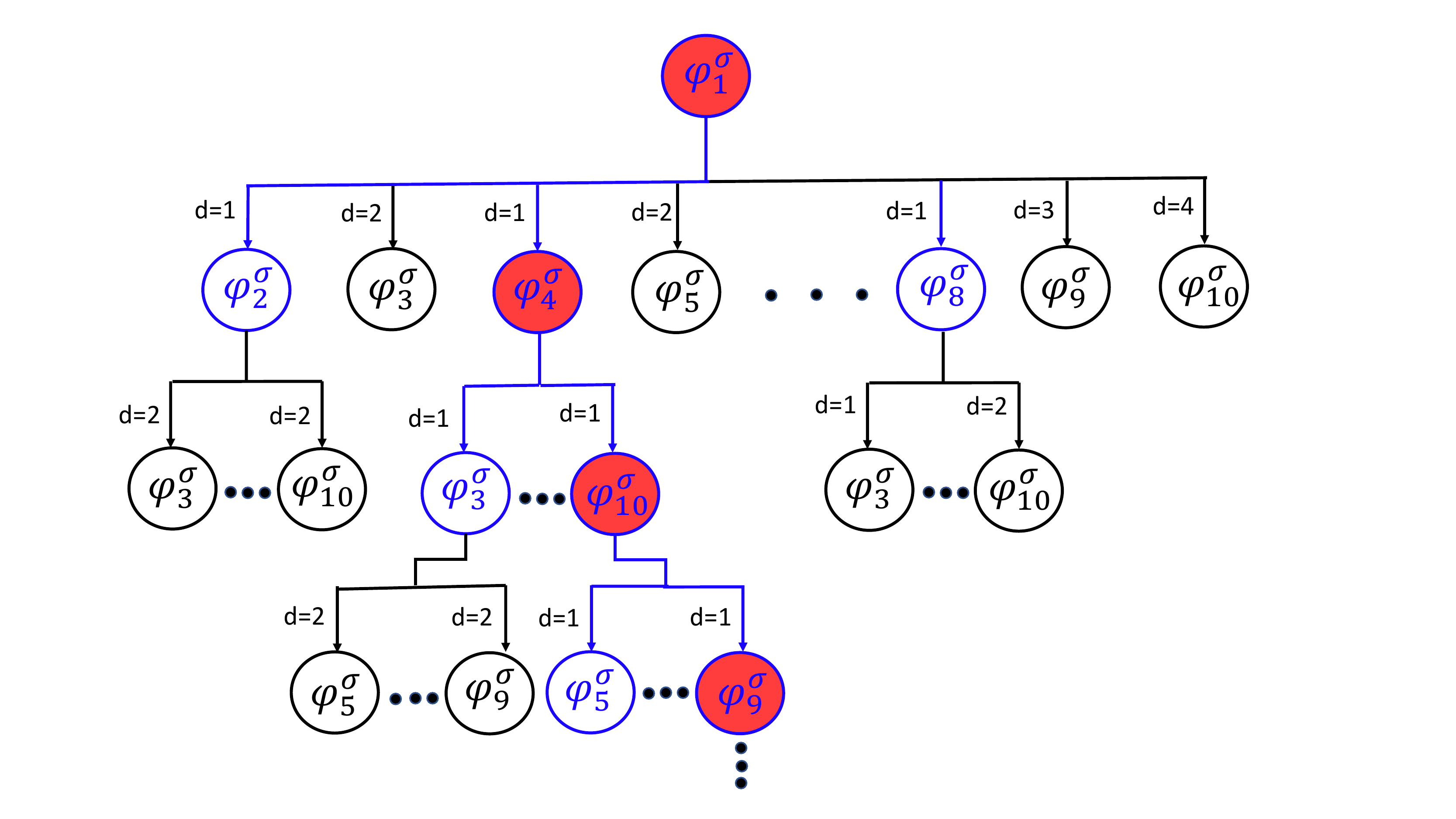}  
\end{center}
\caption{\label{fig:tree} A tree structure of the common 
orbitals $\varphi$, for $10$ Slater determinants. The parent-determinants are filled red circle and the corresponding 
immediate child-determinants are connected by blue lines. 
The distances between a parent determinant and all descendants are labelled 
as $d$, by the
side of the arrows.
The tree structure finds a good sequence of $10$ Slater determinants, which is connected by the blue lines.}
\end{figure}

To use the common
orbitals $\varphi$, we define a tree structure that minimizes the distance between two Slater determinants, where ``distance'' means 
the number of different orbitals between the two Slater determinants. An example of the tree structure for $10$ Slater determinants is shown in Fig.~\ref{fig:tree}.

 We define the first parent of the tree structure, which is the determinant with the largest $|c_m|^2$. 
 In Fig.~(\ref{fig:tree}), the first parent is $\varphi_1^\sigma$ in the filled red circle. Then we calculate the distance between the first parent-determinant and all other determinants, and choose its child-determinants with the shortest distances. 
 In the figure $\varphi_2^\sigma$ ,$\varphi_4^\sigma$, and $\varphi_8^\sigma$ are child-determinants, since they have the shortest distance $d=1$. The second parent is chosen among them. 
We calculate the distances between each candidate and the rest of the determinants. 
The one with the shortest distance is selected as the next parent-determinant.
If two candidates have the same shortest distance, the one with more child-determinants is selected as the next parent-determinant.
For example, $\varphi_2^\sigma$ cannot be a parent determinant since its shortest distance ($d=2$) is larger than that of $\varphi_4^\sigma$ and $\varphi_8^\sigma$  ($d=1$). 
As $\varphi_4^\sigma$ has two child determinants ($\varphi_3^\sigma$, $\varphi_{10}^\sigma$) with $d=1$, while $\varphi_8^\sigma$ has one child-determinant ($\varphi_3^\sigma$) with $d=1$, $\varphi_4^\sigma$ is preferred as a parent determinant.
The same procedure is applied to find the next parent-determinant, 
until all Slater determinants are on the tree structure. 
 
 With the tree structure established, we will only 
 do a full computation 
 for the first parent-determinant. The Sherman–Morrison formula is then used to achieve 
 fast updates for the 
 child-determinants. For the example in Fig.~\ref{fig:tree}, the order during our simulation is:
 \begin{enumerate}
     \item Full calculation on $\varphi_1^\sigma$.
     \item Use the information on $\varphi_1^\sigma$ to calculate $\varphi_2^\sigma$, $\varphi_4^\sigma$, $\varphi_8^\sigma$.
     \item Use the information on $\varphi_4^\sigma$ to calculate $\varphi_3^\sigma$, $\varphi_{10}^\sigma$.
     \item Use the information on $\varphi_{10}^\sigma$ to calculate $\varphi_5^\sigma$, $\varphi_9^\sigma$.
     \item $\cdots$
 \end{enumerate}
 
 \subsubsection{Weight}
 The weight with a multi-determinant trial wave function is given by
 \begin{align}
     W_{k} &= w_{k} \langle \Psi_{T} | \psi_{k} \rangle \\
      &= w_{k} \sum_{m=1}^{N_d} c_m^* 
     \det \left[ (\varphi_{m_\uparrow}^\uparrow)^\dagger \psi_k^{\uparrow} \right] 
     \det \left[ (\varphi_{m_\downarrow}^\downarrow)^\dagger \psi_k^{\downarrow} \right] .
 \end{align}
We next consider the computation of the overlap matrix $\varphi_{m_\sigma}^{\sigma\dagger} \Psi_k^\sigma$.
We define a $\Phi_F^\sigma$ matrix that contains all columns in $\varphi_{m_\sigma}^\sigma$. The corresponding 
overlap matrix is
\begin{equation}
    O_F^\sigma = \Phi_F^{\sigma\dagger} \psi_k^\sigma \, .
\end{equation}
For each $\varphi_{m_\sigma}^\sigma$, we select the rows in the $O_F$ matrix to build the overlap matrix. 
If the overlap matrix of a parent-determinant is $A_p^\sigma$, its child-determinant $A_c^\sigma$, with a distance $d_c^\sigma$, has overlap matrix 
\begin{equation}
    A_c^\sigma =  A_p^\sigma + U^\sigma V^\sigma,
\end{equation}
where $V^\sigma$ is a $d_c^\sigma\times N^\sigma$ matrix and $U^\sigma$ is a $N^\sigma \times d_c^\sigma$ matrix. Note that most of the elements in $U^\sigma$ are zero, with the nonzero elements residing only in the rows 
which are different between $A_c$ and $A_p$. Using Sherman–Morrison formula, 
we can update the determinant of $A_c^\sigma$ by 
\begin{equation}
    \frac{\det A_p^\sigma}{\det A_c^\sigma} = \det\left[1 - (1+V (A_p^{\sigma})^{-1}U)^{-1} V (A_p^{\sigma})^{-1} U\right].
\end{equation}
This is similar to the fast updates in AFQMC widely applied in lattice models;
it has also been applied in diffusion MC with multi-determinant 
trial wave functions \cite{Clark-multi-det-DMC-doi:10.1063/1.3665391}. 

When the child-determinant becomes a parent-determinant, its inverse can be calculated by
\begin{equation}
    (A_c^\sigma)^{-1} = (A_p^\sigma)^{-1} -  (A_p^\sigma)^{-1} U (1+V (A_p^{\sigma})^{-1}U)^{-1} V (A_p^\sigma)^{-1} .
\end{equation}
With $(A_p^\sigma)^{-1}$, we can quickly calculate the overlap between the trial wave function and walkers. Since $(A_p^\sigma)^{-1}$ is also updated through the tree structure, we need to periodically re-calculate the inverse from scratch 
to avoid numerical instability.  

 \subsubsection{Force bias}
 The force bias for multi-determinant trial wave function is
 \begin{align}
      F_\gamma & = \sqrt{-\Delta \tau} \frac{ \langle \Psi_T | \hat{L}_\gamma | \psi_k \rangle }{  \langle \Psi_T | \psi_k \rangle } \\
      & = \sqrt{-\Delta \tau} \sum_\sigma \frac{ \sum_{m=1}^{N_d} c_m^* \langle \varphi_{m_\uparrow}^\uparrow| \psi_k^\uparrow \rangle \langle\varphi_{m_\downarrow}^\downarrow | \psi_k^\downarrow \rangle \frac{\langle\varphi_{m_\sigma}^\sigma | \hat{L_\gamma^\sigma}| \psi_k^\sigma \rangle}{\langle\varphi_{m_\sigma}^\sigma | \psi_k^\sigma \rangle}}
      { \sum_{m=1}^{N_d} c_m^* \langle \varphi_{m_\uparrow}^\uparrow| \psi_k^\uparrow \rangle \langle\varphi_{m_\downarrow}^\downarrow | \psi_k^\downarrow \rangle} .
 \end{align}
 The calculation of $c_m^* \langle \varphi_{m_\uparrow}^\uparrow| \psi_k^\uparrow \rangle \langle\varphi_{m_\downarrow}^\downarrow | \psi_k^\downarrow \rangle$ has been discussed in the previous section. We focus on the local measurement 
 \begin{equation}
   \frac{ \langle\varphi_{m_\sigma}^\sigma | \hat{L_\gamma^\sigma}| \psi_k^\sigma \rangle } { \langle\varphi_{m_\sigma}^\sigma | \psi_k^\sigma \rangle } =   \Tr \left[ \left( \varphi_{m\sigma}^{\sigma}\right)^\dagger L^\sigma_{\gamma}  \Theta_k^\sigma \right],
 \end{equation}
with $\Theta_k^\sigma = \psi_k^{\sigma} \left[ \left( \varphi_{m\sigma}^{\sigma}\right)^\dagger \psi_k^{\sigma} \right]^{-1}$. 
Similar to the force bias for single-determinant $|\Psi_T\rangle$, we only calculate $\left(\varphi_{m\sigma}^{\sigma}\right)^\dagger L^\sigma_{\gamma}$ once through the whole AFQMC simulation. In practice, we calculate $\Phi_F^{\sigma\dagger} L^\sigma_{\gamma}$ and save it to memory. $\left(\varphi_{m\sigma}^{\sigma}\right)^\dagger L^\sigma_{\gamma}$ can be constructed from $\Phi_F^{\sigma\dagger} L^\sigma_{\gamma}$ by selecting corresponding rows, which 
dramatically reduces the memory requirement. 
Note that $\Theta_k^\sigma$ can also be updated from the parent-determinant using the Sherman–Morrison formula
\begin{equation}
    \Theta_{kc}^\sigma = \Theta_{kp}^\sigma - \Theta_{kp}^\sigma U (1+V (A_p^{\sigma})^{-1}U)^{-1} V (A_p^\sigma)^{-1}, 
\end{equation}
where  $\Theta_{kc}^\sigma$ is  the targeted child-determinant
and $\Theta_{kp}^\sigma$ is for the parent-determinant.

\begin{figure}[htbp]
\begin{center}
  \includegraphics[width=1.0\textwidth]{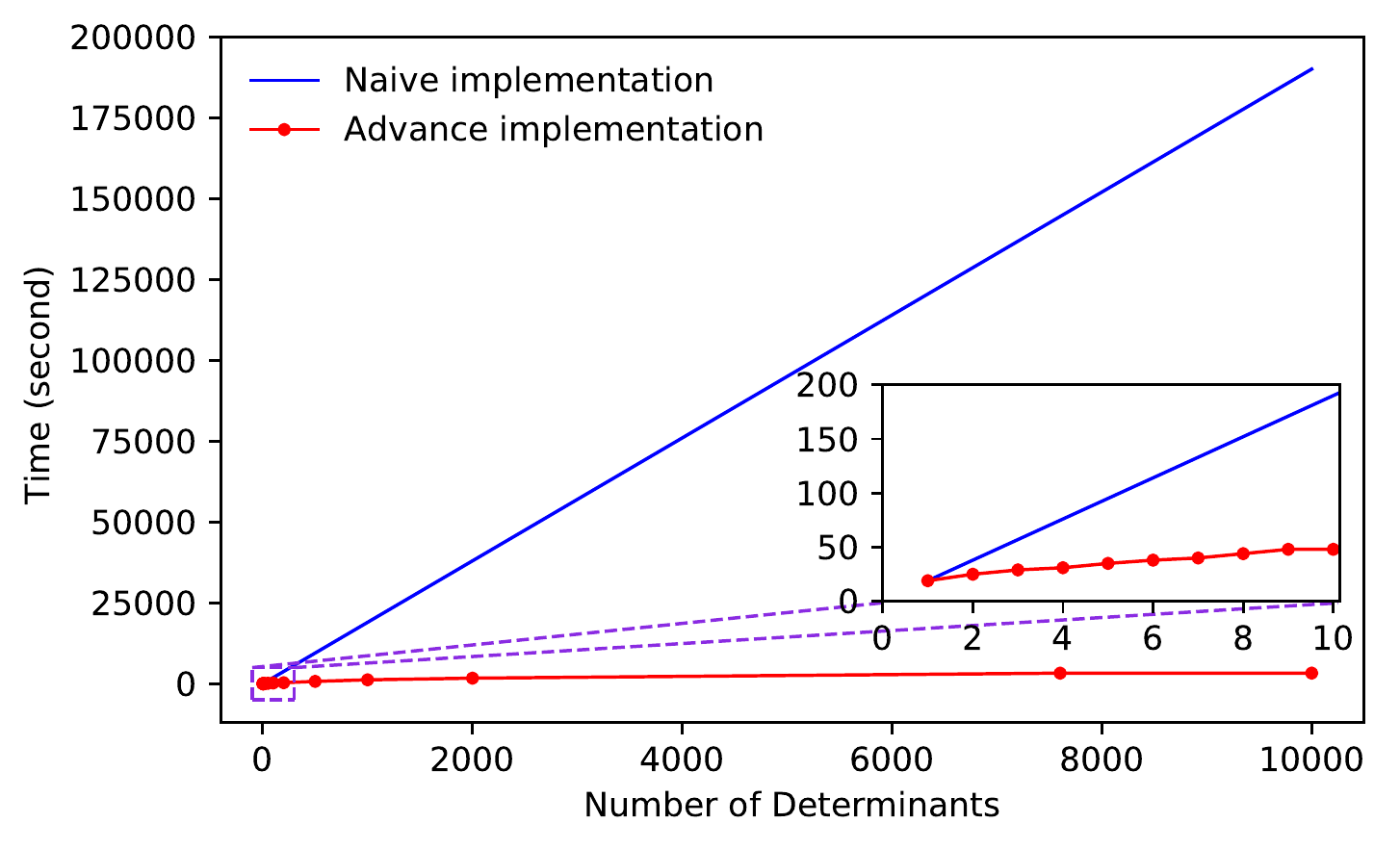}  
\end{center}
\caption{\label{fig:timing} The timing for AFQMC simulations with a multi-determinant trial wave function. The systems is the $O$ atom in the ccpVdZ basis. The naive implementation scales linearly with the number of determinants in the trial wave function, which is shown by the blue extrapolated line. Timing measurements from our implementation is shown by the red dots. With $10,000$ determinants, the speedup is more than $\times 60$.
The inset shows a zoom of up to $10$ determinants in the trial wave function.} 
\end{figure}

 The computation of the interaction 
 energy is similar to the procedure above in 
 computing the force bias, so we omit a more detailed discussion.
 Using the tree structure, the additional scaling with $N_d$ is reduced to sub-linear of $N_d$. A comparison between the naive implementation and 
 the fast algorithm described here is shown in Fig.~\ref{fig:timing}.
 A simple example is used, with multi-determinant trial wave functions from a  CASSCF calculation (obtained with
 PyScf \cite{doi:10.1002/wcms.1340}).
 It is clear that the advanced implementation leads to a drastic speedup. (This implementation 
 was employed in the recent Simons benchmark project \cite{PhysRevX.10.011041}.)
 
 \section{Self-consistent AFQMC}
The trial wave function is used to control the sign and phase problems in the AFQMC calculation. In the previous section, we presented the advanced implementation for 
multi-determinant trial wave functions, which have been shown to systematically yield AFQMC results of chemical accuracy in a large number of molecules (e.g.,~Refs~\cite{PhysRevX.10.011041,Shee-TM-doi:10.1021/acs.jctc.9b00083,Rudshteyn-TM-Complexes-doi:10.1021/acs.jctc.0c00070}). 
The improved scaling of multi-determinant trial wave function 
to sub-linear in $N_d$ thus provides a significant boost towards systematic and general applications in molecular systems. 

In extended systems, however, the number of determinants needed in
the most challenging strongly correlated materials will grow 
grow exponentially 
with system size in a CASSCF-like treatment.
It is thus important to have size-consistent alternatives.
In addition to the single-determinant trial wave function, which has been shown to be very accurate in a large variety of systems, interesting possibilities exist 
with GHF and symmetry-restoration \cite{PhysRevB.89.125129}, BCS \cite{PhysRevA.84.061602}
and HFB \cite{PhysRevB.95.045144}, and a stochastic representation of Jastrow factors \cite{Chang-Jastrow-AFQMC-PhysRevB.94.235144}. Recently 
a self-consistent constraint has been proposed and shown to further reduce the systematic error from the constrained path approximation in lattice model calculations, especially in quantities such as spin and charge density and the reduced density matrix
 \cite{PhysRevB.94.235119,PhysRevB.99.045108}.
Conceptually 
this provides a framework in which  the outcome of one 
AFQMC calculation can be fed into the next iteration to achieve a systematically improvable self-consistent procedure. 


Here we consider the generalization of the 
self-consistent AFQMC idea to \emph{ab initio} computations in molecules and solids. Two flavors of the self-consistent 
approach have been suggested \cite{PhysRevB.94.235119}.
The first is 
to 
couple the AFQMC calculation to an independent-electron calculation. 
\cite{PhysRevB.94.235119, PhysRevB.99.045108}. 
The reduced one-body density matrix obtained from AFQMC is fed back to the independent-electron calculation which can be, for 
example, HF. The effective interaction in the HF is tuned so as to produce a density (or density matrix) which best matches the AFQMC result.  
The output Slater determinant wave function is then used for another AFQMC calculation. The process is iterated until the density matrix is converged. In lattice model calculations, this self-consistent procedure often involved tuning an effective 
$U$ parameter. 
The second flavor of the self-consistent approach is to 
diagonalize the AFQMC density matrix and select natural orbitals up to $N_\sigma$ for the trial wave function, and to iterate until the resulting natural orbitals do not change any further. 

\COMMENTED{
The independent-electron calculation can be an Hartree-Fock or a density-functional theory (DFT) calculation. For example, we tune the effective interaction in the Hartree-Fock to match the density matrix from AFQMC calculation, which provide an optimal effective "U" parameter in Hartree-Fock. Similarly, the self-consistent procedure can also find an optimal functional in the context of DFT. The most simple independent-electron calculation is to diagonalize the density matrix and select natural orbitals up to $N_\sigma$ for the trial wave function. 
}

 \begin{figure}[htbp]
\begin{center}
  \includegraphics[width=0.9\textwidth]{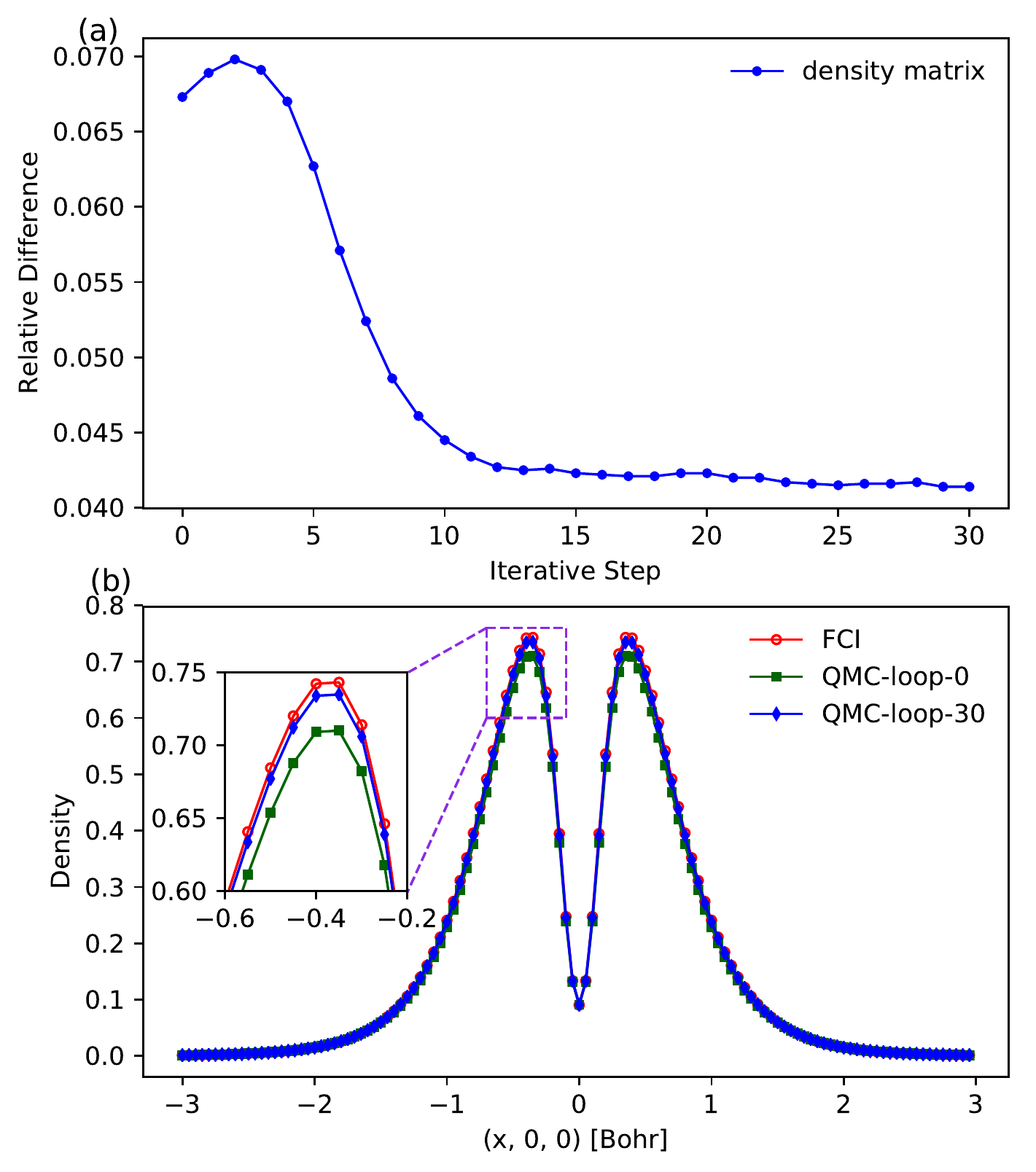}  
\end{center}
\caption{\label{fig:self-consistent} Self-Consistent AFQMC calculation using the natural orbitals of the computed one-body density matrix.
Results are shown for the O$^+$ atom in the cc-pVdZ basis, using an ECP.
(a) The relative difference of one-body density matrix of the majority spin (4 electrons) computed by AFQMC with respect to FCI results, 
$|| G_{\rm AFQMC}-G_{\rm FCI}||/||G_{\rm FCI}||$, as a function of self-consistent iteration steps.
(b) The computed electronic density of the majority spin on a line cut through the atom (distance $x$ in Bohr). The density from AFQMC using 
the HF trial wave function (loop 0) is improved by the self-consistency upon convergence, as compared to the FCI density.
}
\end{figure}

We first study the second approach, self-consistent computation 
from diagonalizing the density matrix, using a simple example. In Fig.~\ref{fig:self-consistent}, results are shown for the O atom
using an effective-core potential (ECP, see Ref.~\cite{PhysRevX.10.011041} for details of the ECP),
with $4\uparrow\,2\downarrow$ electrons.
The AFQMC simulation starts from a HF 
trial wave function. The density matrices 
calculated by the AFQMC simulation using this trial wave function 
are diagonalized for both spin species, and the resulting 
$N^\sigma$ natural orbitals are taken 
to form a single Slater determinant wave function, which is then used as a trial wave function in the new AFQMC simulation. The procedure is iterated until the density matrices are converged. 
Interestingly, starting from a HF trial wave function, the majority-spin density matrix shown in the top panel exhibits 
a non-monotonic behavior, which converges after $15$  iterations.
We see that both the density matrix and the electronic density are improved upon convergence of the self-consistent loop.

We next generalize the first self-consistent method mentioned above to the formalism for real materials, coupling AFQMC to an effective independent-electron calculation.
In particular, rather than viewing a DFT (including HF) calculation as an independent-electron method to treat the \emph{original} Hamiltonian, we view it as a proxy calculation with an \emph{effective} Hamiltonian whose goal is to produce a single determinant with an electronic density (or density matrix) best matching that of the AFQMC (always done with the original Hamiltonian, of course). One way to think about the effective Hamiltonian 
is in terms of the many different DFT functionals in existence, as well as additional ones in which the strength of the Coulomb interaction is allowed to deviate from the true electron-electron interaction. In this way, the self-consistent AFQMC procedure, in addition to being a systematically 
improvable many-body method via the trial wave function, can also  be viewed as an automatic ``screener'' for density functionals.  

\begin{figure}[htbp]
\begin{center}
  \includegraphics[width=0.9\textwidth]{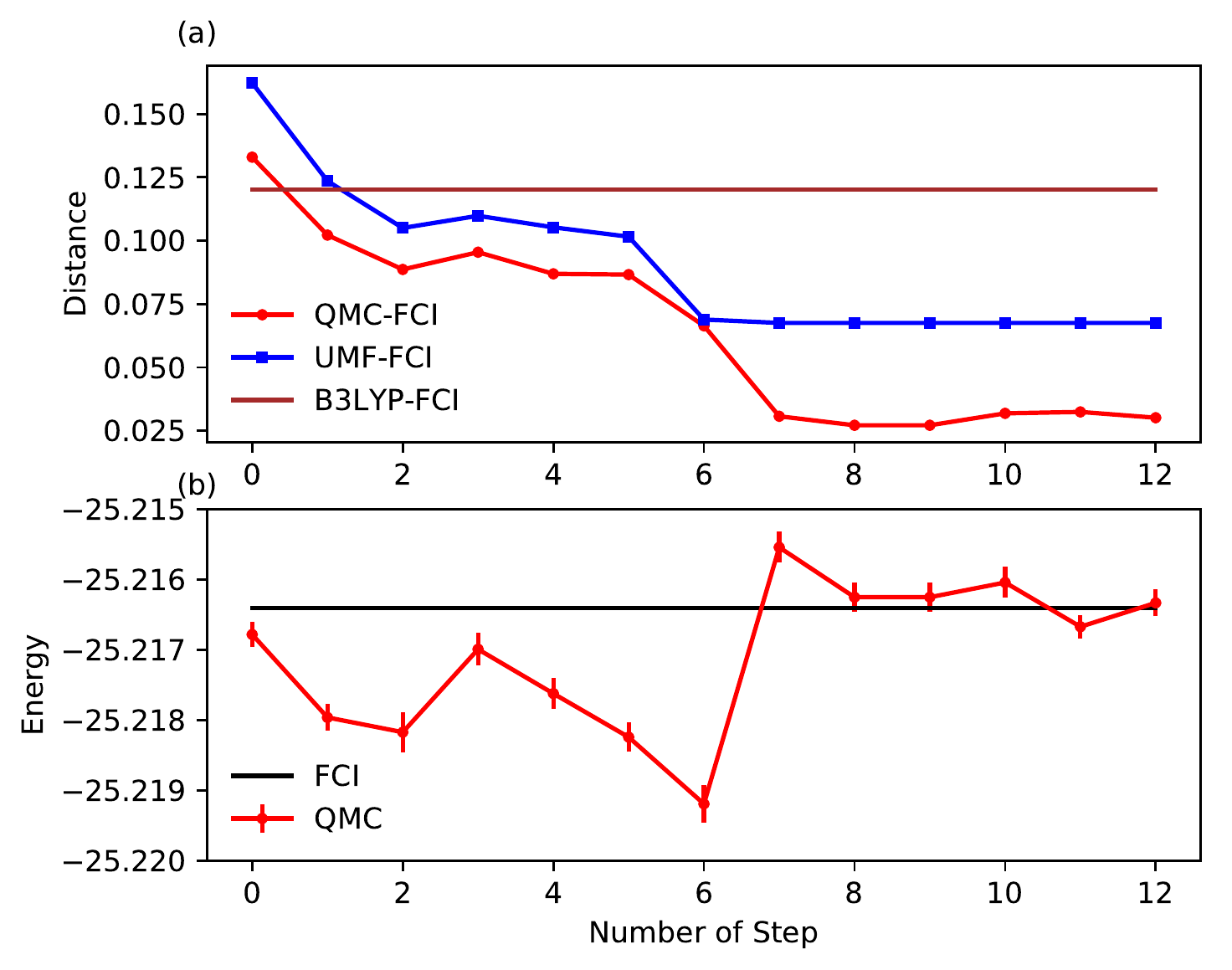}  
\end{center}
\caption{\label{fig:self-consistent-bh}
Self-Consistent AFQMC optimization of an  effective DFT Hamiltonian. 
The AFQMC calculation is coupled to an independent-electron calculation using an effective Hamiltonian resembling the B3LYP functional. The independent-electron calculation produces a single-determinant trial wave function for AFQMC, from which the computed 
density matrix is used to find a new effective Hamiltonian. Results
are shown for the BH molecule in the ccpVdZ basis, versus self-consistent iteration steps. (a) The relative discrepancy of the computed one-body density matrices.  
The self-consistent iteration converges after $8$ steps. 
Note that ``DFT'' (UMF) results also improve during the iteration. (b)  AFQMC energy versus iteration. 
}
\end{figure}

We  illustrate this approach using the BH molecule as an example. 
For the auxiliary  independent-electron calculations, we use  a form for the effective Hamiltonian closely resembling the popular B3LYP functional (as implemented in Pyscf \cite{doi:10.1002/wcms.1340}):
\begin{equation}
    E_{xc} = \beta \left[\alpha\,E_x^{\rm HF} + (0.1-0.1\,\alpha) E_x^{\rm LDA} + (0.9-0.9\,\alpha) E_x^{\rm B88}+(1-d)\,E_c^{\rm LYP} +d\,E_c^{\rm VWN} \right]\,.
\end{equation}
In B3LYP, $\beta=1.0$,  $\alpha=0.2$, and $d=0.19$.
The parameter $\alpha$  tunes the percentage of the exact exchange, 
and $\beta$ scales the effective strength of the Coulomb interaction.
We will allow both parameters to vary in our auxiliary  independent-electron calculations. 
We start our self-consistent process from an initial trial wave function generated 
with the parameters $\alpha=0.80$ and $\beta=1.0$,  which is way off from B3LYP or any reasonable mean-field approximation.
This yields a ground-state energy which is rather accurate at step 0
but with poor AFQMC result on the one-body density matrix, as 
shown in Fig.~\ref{fig:self-consistent-bh}. 
We then perform our auxiliary ``DFT'' calculations by
tuning $\alpha$ and $\beta$ with an interval $0.005$, identifying the 
parameter choices which
minimizes the difference between the DFT density 
matrix and that from AFQMC in the previous iteration. 
The resulting DFT wave function  is fed 
into the next step AFQMC as trial wave function. This process 
reduces the density matrix bias (while giving non-monotonic results 
in the AFQMC total energy from the mixed estimator, which is not variational \cite{Zhang_Book_2019,PhysRevLett.90.136401}), and converges to $\alpha=0.35$ and $\beta=0.975$.
This is far away from the initial parameter choices, and reasonably close to the B3LYP values. 
As seen in the top panel, the DFT density matrix result also improves with the AFQMC, 
and the final answer at the selected parameters is in fact better than that from B3LYP. 
It is worth noting that these results are for 
a finite basis set.
DFT functionals including B3LYP
are designed for the complete 
basis set limit, and there can be non-negligible effects in 
comparing them.
However, this is not relevant to  the point of our test, which 
shows that the self-consistent procedure can find an effective 
Hamiltonian which yields a better description of the 
particular Hamiltonian.

This simple example serves as a proof-of-concept demonstration of the coupling between AFQMC and an auxiliary independent-electron calculation to achieve self-consistency, matching the density matrix (or the electronic density). Clearly the procedure can be made more general and elaborate.
For example, for simplicity we did not vary the correlation part of the functional, 
other than the overall scale $\beta$. 
One could introduce many more parameters, and choose any physically motivated form of the effective Hamiltonian. The results above
show that this is a very promising avenue for not only systematically improvable AFQMC calculations in real materials but also screening DFT functionals which could be 
used in related and larger systems, or be coupled to AFQMC for embedding \cite{Virgus-Co-graphene-PhysRevLett.113.175502} to 
extend system size.

\COMMENTED{expand discussion on DFT/HF a bit more. HF can be viewed as a special case of DFT. The point above is to say we
can tune the functional self-consistently, for example, 
tune the exact-exchange percentage - do we have any results on this?}

\COMMENTED{a 2nd, related point is the self-consistent procedure can be used as a functional 'screener' or 'selector'. We 
compare the DM produced by different DFT with AFQMC, and see
which one is closer. Half-baked but I will think about this more}

\section{Summary}
In this article, we reported our recent progress on several fronts in continuing to develop  the AFQMC method for real materials.
The AFQMC method is highly accurate for a wide range of systems, as demonstrated by recent benchmark studies \cite{Zheng1155,PhysRevX.5.041041,PhysRevX.7.031059, PhysRevX.10.011041, 2019arXiv191101618M}. It has a low-power scaling with the size of the systems and is naturally parallel on high performance computing platforms. The AFQMC method can be applied to any Hamiltonian which can be written in the MC Hamiltonian form, with general one-body and two-body interaction terms. We also presented 
details of the method in a way that facilitates efficient implementations, including advanced implementations which dramatically speed up the algorithm and reduce memory cost. We proposed the use of self-consistent constraints in molecules and solids, 
and studied the behavior under two different flavors.  

The AFQMC method is very promising as a general computational method for 
strongly-correlated many electron systems. Many directions can be pursued in its development and application, for example, further reducing the scaling and improving the efficiency of the algorithm, computation of observables including 
imaginary-time correlations, 
finite-temperature AFQMC for materials and and excited state calculations, etc.  With increased attention and effort in the development of AFQMC both algorithmically and in software, many more applications can be expected in general material systems.  

\section{Acknowledgements}
We acknowledge helpful discussions with Mario Motta, Wirawan Purwanto, and Mingpu Qin.
The Flatiron Institute is a division of the Simons Foundation. This work was conducted using computational resources and services at the Flatiron Institute. 

\section{Data Availability}

The data that support the findings of this study are available from the corresponding author upon reasonable request.


%
%

%




\bibliography{main.bib}

\end{document}